\documentclass[%
reprint,
nofootinbib,
 amsmath,amssymb,
 aps,
]{revtex4-2}

\usepackage{graphicx}% Include figure files
\usepackage{dcolumn}% Align table columns on decimal point
\usepackage{bm}% bold math
\begin{document}

%\preprint{APS/123-QED}

\title{Proposal for a distributed, community-driven academic publishing system}

\author{Matteo Barbone}
\email{matteo.barbone@wsi.tum.de}
\affiliation{%
Walter Schottky Institut, TU M\"unchen, Am Coulombwall 4, 85748 Garching, Germany}%
\affiliation{%
Munich Center for Quantum Science and Technology, Schellingstr. 4, 80799, Munich, Germany}%

\author{Mustafa Gündoğan}%
\email{mustafa.guendogan@physik.hu-berlin.de}
\affiliation{%
Institut für Physik, Humboldt Universität zu Berlin, Newtonstr. 15, 12489 Berlin, Germany}%

\author{Dhiren M. Kara}
%\email{}
\affiliation{%
Cavendish Laboratory, University of Cambridge, JJ Thomson Avenue, CB3 0HE Cambridge, UK
}%

\author{Benjamin Pingault}
\email{bpingault@seas.harvard.edu}
\affiliation{%
John A. Paulson School of Engineering and Applied Sciences,
Harvard University, Cambridge, Massachusetts 02138, USA
}%
\affiliation{%
QuTech and Kavli Institute of Nanoscience, Delft University of Technology, 2600 GA Delft, The Netherlands
}%

\author{Alejandro Rodriguez-Pardo Montblanch}
\email{ar820@cantab.ac.uk}
\affiliation{
QuTech and Kavli Institute of Nanoscience, Delft University of Technology, 2600 GA Delft, The Netherlands
}%

\author{Lucio Stefan}
\email{lucio.stefan@nbi.ku.dk}
\affiliation{%
Center for Hybrid Quantum Networks, Niels Bohr Institute, University of Copenhagen, Blegdamsvej 17,
2100, Copenhagen, Denmark
}%

\author{Anthony K. C. Tan}
\email{a.tan@imperial.ac.uk}
\affiliation{
Blackett Laboratory, Imperial College London,  SW7 2AZ London, UK
}%

\author{\textit{Authors are listed in alphabetical order}}

\date{23rd of April, 2023}

\begin{abstract}
We propose an academic publishing system where research papers are stored in a network of data centres owned by university libraries and research institutions, and are interfaced with the academic community through a website. In our system, the editor is replaced by an initial adjusted community-wide evaluation, the standard peer-review is accompanied by a post-publication open-ended and community-wide review process, aiming at a more objective and longer-term evaluation, the publishing costs are reduced to the running costs of the servers, and access is fully open.
Our proposal addresses the fundamental problems of the current system: it reduces publishing costs, allowing easier access by less well-funded institutions (especially from developing countries); it makes the editorial evaluation distributed and more transparent; it speeds up the peer review process by eliminating the need for multiple resubmissions; and it introduces a long-term, community-wide evaluation of papers, ensuring their continued relevance and accuracy; while maximising its main goals, i.e. ensuring the highest quality of peer review and giving the best referees, the most visibility and the most credit to the best papers.
Our scheme is time-efficient, financially sustainable, ethically fair and represents a significant improvement over the current system.
\end{abstract}


%\tableofcontents

\maketitle

\section{\label{current}Introduction}
Current academic dissemination is based on journals printed by companies and scientific societies specialized in research publishing: researchers summarize their findings in academic papers, which are sent to specialized journals for publication, where they are subjected to the judgement of editors and peer academics. If a paper is accepted for publication, the authors are required to pay a publication fee covering part of the publication costs. If the paper is rejected, academics will re-submit it to a different, usually less prestigious journal, and in an iterative way repeat the process until acceptance. Once published, the access to the paper is, with exceptions in the case of “open access” journals, subject to the payment of a subscription fee to the publisher.\\
The current system suffers from a number of shortcomings. Publication fees are very costly; journals require expensive subscriptions; the first evaluation of a manuscript is carried out by one or few editors that may not be specialists of the field, but need to quickly take decisive choices for the future impact of a paper; the peer review process typically takes many months to complete, but despite its length, it is still subject to time and referee constraints that prevent it from providing a comprehensive view of the relevance, accuracy and reproducibility of a finding, and thus from acting as a solid barrier against malpractice.

\subsection{\label{commercial}Commercial academic publishing}
The current academic publishing system is today widely debated, with questions being raised on its impact on the academic community, on the progress of knowledge, and on society at large\footnote{\label{1}http://thecostofknowledge.com}$^{,}$\footnote{\label{2}https://www.theguardian.com/science/2012/apr/09\\/frustrated-blogpost-boycott-scientific-journals}$^{,}$\footnote{\label{3}https://www.theguardian.com/science/2017/jun/27\\/profitable-business-scientific-publishing-bad-for-science}. 
Currently, academic publishing is worth ~£19 billion a year\footnotemark[3], constituting a relevant source of expenses for institutions, which often struggle to pay for access to research articles and negotiate lower fees. A regime of oligopoly, where a handful of publishers dominate the market, and unrewarded peer reviewing, which requires a high degree of specialised skills and time commitment and is at the core of the scientific evaluation, make scientific publishing most profitable. To put that in perspective, major scientific publishers have profit margins above 30\%\footnote{\label{4}https://www.ft.com/content/893252b8-31c3-11e8-b5bf-23cb17fd1498}$^{,}$\footnote{\label{5}https://www.relx.com/$\sim$/media/Files/R/RELX-Group/documents/reports/annual-reports/relx-2021-annual-report.pdf}, superior to those of tech companies such as Alphabet\footnote{\label{6}https://abc.xyz/investor/static/pdf/2021Q4\_alphabet\\$\_$earnings$\_$release.pdf?cache=d72fc76}.
This comes at the cost of customers, namely research institutions, who struggle to pay the ever-increasing subscription fees\footnote{\label{7}https://www.theguardian.com/science/2012/apr/24/harvard-university-journal-publishers-prices}, researchers, who spend a significant portion of their research budgets to publish, and the general public, who is denied free access to research results usually paid through taxes. \\Open-access, where researchers pay upfront hefty fees\footnote{\label{8}normally 1000-5000 EUR per paper; a list is available at https://www.springernature.com/gp/open-research/journals-books/journals} to make the research free for all readers, does not solve this problem either, as costs are shifted but not dropped\footnote{\label{9}https://www.ft.com/content/8dc9c370-492d-11e8-8ae9-4b5ddcca99b3}, and many institutions cannot bear them.\\
In the past few years, a plethora of novel fully open-access and peer-reviewed journals were founded. Such efforts highlight the need for a more diversified and community-based approach to scientific publishing, but do not tackle all of its other limitations, elaborated below.
%\subsubsection{Wide text (A level-3 head)}
%The \texttt{widetext} environment will make the text the width of the
%full page, as on page~\pageref{eq:wideeq}. (Note the use the
%\verb+\pageref{#1}+ command to refer to the page number.) 
%\paragraph{Note (Fourth-level head is run in)}
%The width-changing commands only take effect in two-column formatting. There is no effect if text is in a single column.

\subsection{\label{editors}Editors}
Journals are not all equally prestigious, and space in research journals, both online only or printed, is limited. Researchers are thus constantly fighting to publish in the most prestigious journals, where their work can receive the best level of peer-reviewing, the most visibility in the community, the most prestige, and the most citations.
The careers of researchers, especially at the earlier stages, are bonded to their publication output. Publishing in high-impact journals represents the path towards higher chances of obtaining faculty positions and research grants from funding agencies. 
Once a manuscript is submitted for publication, an editor judges its potential impact on the research community. Some journals reject 80\% of the submitted papers at the initial editor’s screening\footnote{\label{10}http://www.sciencemag.org/site/feature/contribinfo/faq/\\index.xhtml}. If the editors find a paper suitable for their journal, they ask to other experts (called referees), who offer their service for free, to judge the technical correctness and novelty of the paper under examination (peer-review system). Based on the reports of the referees, editors decide whether to accept or reject a manuscript.\\
As they can reject a manuscript before peer-reviewing, they choose the referees, they judge referee reports, and they take the ultimate decisions on whether to publish or not each article, editors hold significant power. It is thus in the interest of the community that editors be as much as possible independent, objective and have a deep understanding of the research field they deal with. These are all very difficult requirements to meet. An editor may have a personal connection with an academic, be influenced by their prior achievements, have an incomplete understanding of a research topic, or be subject to following precise editorial guidelines conflicting with the interests of the academic community and the society.\\
Thus, subjective editorial choices are a crossroad in the life of researchers.

\subsection{\label{referees}Referees}
Referees are early career researchers, academics or former academics who evaluate the impact and technical correctness of papers, offering their expertise to the community for free. Each paper is usually refereed by 1-3 referees within a limited time-frame of a few weeks.\\
On one side, due to the current incentive scheme, academics will push to have their findings published in the highest impact journals possible, which may result in the same manuscript being submitted and reviewed multiple times, overall involving a large number of referees and multiplying the load on the community. On the other side, analogous to editors, also referees should be independent, objective, and possess a strong familiarity with the research topic under scrutiny. However, referees may miss key technical errors that cast serious doubts on a paper's findings, may under- or over-estimate the relevance of a paper for its community, etc. Thus, refereeing is a steep challenge, and expecting a flawless result would be ungenerous. \\

Importantly though, with the constantly rising pressure on academics to publish much and high impact, and the consequent increased risk of shortcuts, lack of rigour, mistakes, and malpractices, the best possible combination of time-efficiency and highest level of peer-review available is not only crucial for the career of academics, but is required more than ever as the strongest safeguard against unfounded claims, non-reproducible research, and fraud. \\
Desirable steps would include expanding the evaluation by increasing the referee base in number and time, acknowledging the work of referees, and provide fairer exposure to papers by limiting potential biases tied to the journal in which a given paper is published.\\
Scientific publishing can take advantage of the power of worldwide accessibility of data and near-instant communications to involve much more efficiently a larger portion of the community to directly and readily participate in the evaluation of its results.

\section{\label{structure}Structure of the proposal}
\subsection{\label{portal}Online portal}
\begin{figure*}
\centering
\includegraphics[width=0.92\textwidth]{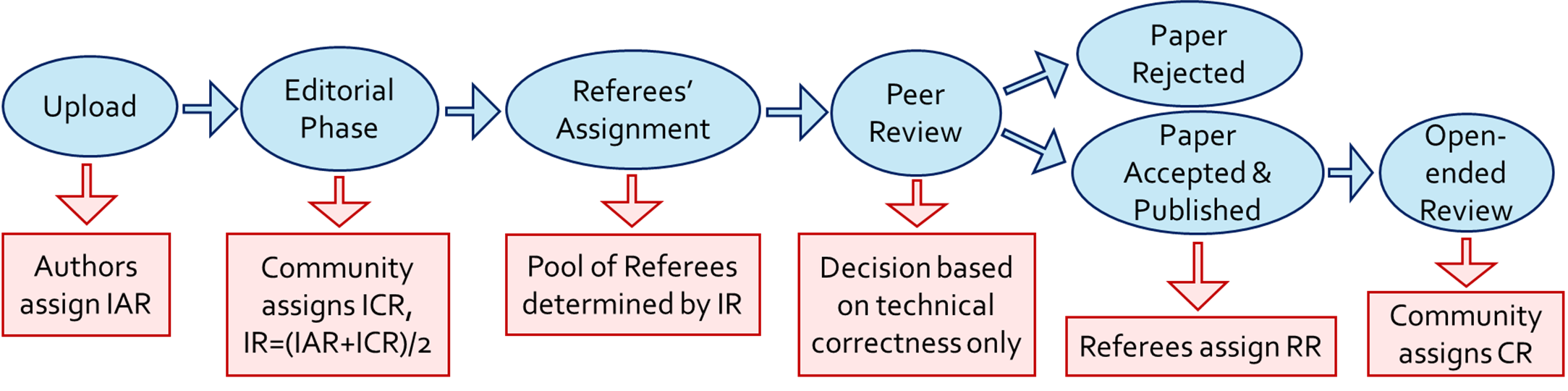}
\caption{\textbf{Workflow of the publication and evaluation process according to the proposed publishing system}}
\label{fig:Figure1}
\end{figure*}

We propose a scholarly publishing system centred on and led directly by the broadest representation of the academic community, and built on a distributed network of servers hosted by university libraries and research institutes, where articles are stored. \\An online portal connects the servers and provides the necessary infrastructure for uploading, peer reviewing, searching and disseminating documents Each user registers on the portal and creates an account with an institutional and verified email address or analogous (i.e. ORCID), similarly to Google Scholar or arXiv today, adding their fields and sub fields of expertise.\\
To structure the visibility of the papers in a hierarchical and topic-related way, the main online portal branches out in sub portals, hosting more and more specialized papers. As an example, the parent domain (e.g. the hypothetical domain www.scientia.org) features the papers suitable for the most general audience (i.e. ``higher impact''). Sub domains (e.g. www.scientia.org/physics,  www.scientia.org/biology...) and sections arranged by keywords feature specialized papers and differ according to the topic.
Users who register will specify their field and sub fields. \\
Next, we detail the manuscripts' submission, evaluation, and publication process. Fig. \ref{fig:Figure1} summarizes the complete workflow.

\subsection{\label{uploading}Manuscript uploading and evaluation}
Each registered user can upload a fixed number of manuscripts per year, to encourage quality submissions over quantity, indicating the keywords to which the work belongs to. Once submitted, the article will be visible online for a limited period of time (editorial period), e.g. 1 week, in a section of the portal dedicated to new submissions. The authors will be asked to give an initial self-assessment of the impact of their work and the suitability for a wide audience (Initial Authors' Rating -IAR).

\subsubsection{\label{community}Community-wide editorial phase}
The editorial phase occurs as a collaborative community effort during the cool down period. Each registered user whose field overlaps of a determined amount with the topic and subtopics of the manuscript will be entitled to rate the impact of the manuscript. The higher the IAR, the smaller the overlap required (so the larger the participation to the initial evaluation). Thus, a higher IAR also exposes a paper to the scrutiny of a wider audience. The average community rating will form the Initial Community Rating (ICR). To promote fairness of judgement, tails in the community rating distribution (e.g. highest and lowest 20\% ) may be excluded.\\
At the end of the editorial phase, an Initial Rating (IR) will be formed as average of the IAR and of the ICR. The IR will be used to restrict the pool of reviewers (for details on the referees' assignment see section~\ref{rating}) among the users with compatible expertise with the keywords of the manuscript. 

\subsubsection{Standard peer-review and referees' evaluation}
Once the referees are assigned, the IR is deleted, and peer-review starts, conducted as it is today as private correspondence between the authors and the referees. Each part will be able to choose anonymity or reveal their names. Referees will be asked to judge the technical correctness of a paper, and to give a rating to the paper.\\
Referees may be asked, before accepting the work, to specify a required timeline (e.g. 2-4 weeks) to complete the refereeing process. To prevent systematic cases of delay, the number of referees may be increased by one more than the minimum necessary number, thus a late referee may be automatically excluded in order to guarantee a timely refereeing process.\\
If the paper is accepted for publication, the referees will also give a Referee Rating (RR) on the novelty/general interest level of the paper, to judge the impact. The RR will be the starting point to determine the visibility of the paper on the online portal. The higher the RR, the higher the visibility of the paper. \\
If referees reject the publication due to serious structural flaws, all co-authors may be required to wait a fixed amount of time (e.g. 3 months) before uploading any new manuscript. This will prevent the authors from immediately trying their luck with new referees.\\
After peer-reviewing and acceptance, the paper will be added to the list of peer-reviewed papers (published online). The names and affiliations of the authors will appear in full, and a final comment from the reviewers as well as their names may also be included. \\

We underline the time-efficiency of the scheme, as it eliminates the habitual protocol of submitting to, and undergoing, peer-review in multiple journals, starting with the highest impact ones all the way down following rejections, which  unnecessarily stretches the review length and requires unnecessarily long time and many referees. On the contrary, due to the single submission process, more referees can be involved while keeping the average number of referees per paper significantly lower than today. For instance, today, a paper being submitted three times before being accepted for publication usually would involve three or more editors and three to nine referees, and would  require at each new submission about a week of editorial evaluation and a month or more before referees complete each round of review, for a total minimum of four months. A process requiring a single submission involving 4-5 referees would reduce the number of total referees involved by up to about a half, while around doubling their minimum number, and cutting the waiting time by months. The larger number of referees may also be advantageously used to remove the extremes when assigning the RR, to encourage a fair assessment. For instance, if five referees were appointed, and only four submitted a report on time, forming a RR from the two middle ratings would be equivalent to what commonly happens today when two referees review a manuscript submitted to a journal with defined impact factor, and both suggest to accept it.

\subsubsection{Open-ended community-wide review}
Post peer-review, once the paper is published and accessible in its final form, verified users will be allowed to rate again the paper with a permanent Community Rating (CR) but may be required to add a non-void comment to promote a thorough read, and their name may appear publicly, to promote fairness of judgement and transparency. Also in this case, tails in the rating distributions may be removed when calculating the CR.\\
The paper will thus have 2 ratings: the RR from the referees, immutable, and the CR from the community, open-ended as more users in time will read, rate and comment the paper. The combination of both ratings, the evolution of the CR rating in time, and the citations will offer richer metrics and thus more nuanced evaluation of each work compared to what is possible today. \\
Moreover, the combination of referees and users significantly eases the problem of appeals: if the authors were not to be satisfied with the RR, they will have the chance to compare their opinion with the wider community. Further, if a paper were to be fully or partly irreproducible, or concerns regarding its rigour were to arise, the CR would progressively trend lower, thus raising a red flag for the community.
%Also, correlation patterns in the RR attributed by a referee or the CR attributed by a user compared for instance to the average CR may be used 

\subsubsection{\label{rating}Insight on the referees' assignment process}

Referees are chosen by an algorithm among the users with a tight overlap of expertise with the keywords of the manuscript, a predetermined minimum number of total published papers and a minimum number of published papers of specific novelty rate. The higher the IR, the higher the number of total papers, plus the higher number of papers above a post-publication rating (explained later), the referee should possess. A possibility is shown in Table~\ref{t1}.
The authors are still allowed to exclude specific referees and suggest others. The algorithm may be designed to directly include one of the suggested referees or weigh more their chances of being selected. A regular and significant discrepancy between IAR and ICR may be used to find authors who consistently overestimate or oversell the impact of their works, or users who consistently under or over vote papers of specific authors, and forms of discouragement can be designed to limit such practices, such as a temporary upper limit on their IAR, or a lower number of available uploads.\\
\begin{table}[b]%The best place to locate the table environment is directly after its first reference in text
\caption{\label{t1}{Possible bibliometric requirements for referees according to the IR of papers after the editorial phase.}
}
\begin{ruledtabular}
\begin{tabular}{cccc}
\multicolumn{1}{c}{IR}&\multicolumn{1}{c}{Min. Number}&\multicolumn{1}{c}{Min. Number}&…of RR\\
\multicolumn{1}{c}{}&\multicolumn{1}{c}{of papers}&\multicolumn{1}{c}{of papers}&or CR\\
%\textrm{of papers…}
\colrule
\multicolumn{1}{c}{1-2} & 3 & 2 & 2\\
\multicolumn{1}{c}{2.01-3} & 5 & 2 & 3\\
\multicolumn{1}{c}{3.01-4} & 8 & 1 & 4\\
\multicolumn{1}{c}{4.01-5} & 10 & 2 & 4\\
\end{tabular}
\end{ruledtabular}
\end{table}

\subsubsection{\label{rating}Insight on the rating scale}
As we have seen, each paper receives multiple ratings by multiple bodies, which serve the purposes that today are mainly performed only by the impact factor of journals.
%: selecting the reviewers, identifying the broadness of readership, determining the visibility of each manuscript on the portal, unlocking the community and sub-communities allowed to participate in the post-publication peer-review. \\
%For instance, it allows to implement algorithms that highlight, in the various home pages, the highest rated papers within a certain timeframe, thus ensuring visibility in a transparent way.  \\
A possible rating scale is as follows:\\

\noindent
$\mathbf{5}$: ground-breaking advancement, of the broadest cross-disciplinary interest (e.g top 5\%).\\
$\mathbf{4}$: very significant advancement, of general interest for a whole discipline or closely related disciplines (e.g. top 15\%, for all physicists).\\
$\mathbf{3}$: quite relevant advancement, of interest for a subdiscipline (e.g. top 30\%, for all condensed matter physicists). \\
$\mathbf{2}$: specialized finding, of interest for a branch of a subdiscipline (e.g. top 50\%, only for condensed matter physicists working on layered materials).\\
$\mathbf{1}$: very specialized finding, of interest for a specific branch of a subdiscipline (e.g. outside top 50\%, only for condensed matter physicists working on layered ferromagnetic materials).\\
$\mathbf{0}$: (only available as CR): doesn't meet minimum requirements of scientific rigour, serious concerns on reported conclusions.\\

To prevent inadequate referees or users trying to inappropriately influence the RR or CR, and promote fairness of judgement, we have suggested approaches such as eliminating extremes in the rating distribution. However, more proactive solutions may also be considered, such as analyzing correlation patterns between the RR attributed by a referee or the CR attributed by a user to manuscripts of specific authors compared for instance to the average CR, or concealing the authors' names during the editorial phase and/or the peer-review.

\subsection{Governance and editorial format}

\subsubsection{Governance}

A key aspect of a distributed system is the central role played by the procedures for dealing with improvements and changes, both quantitative, such as adjusting the maximum number of uploads or the requirements for reviewing papers, and qualitative, such as changing the architecture of the system, introducing new sections, evaluation methods, ratings, etc. 
The platform will have a dedicated section where any member of the community can propose new changes, which will first be discussed among users in a forum-like setting, then voted on and implemented if approved, similar to what happens today in the Quantum journal, for example.\\
We envisage the possible presence, outside the regular peer review process, of a limited number of curators, elected or appointed by the institutions that make up the nodes of the network, who would be responsible for managing the financial and technical stability of the platform, overseeing the implementation of approved changes, handle unforeseen necessities, bringing complex controversies to the community, possibly commissioning reviews, etc.

\subsubsection{Editorial format}

The portal should be open to all content type analogous to what is currently done: articles/ letters, reviews, possibly even books.\\
Guidelines for paper formatting could reflect the initial novelty of the paper and the type of article. Papers will receive a DOI and will continue to be cited as they are today.

\subsection{Rewards for referees}
Today, refereeing is a free service that academics give to the community by dedicating 68.5 million hours per year to peer-reviewing\footnote{\label{11}The Global State of peer-review report 2013-2017}. If this was to be payed at an hourly rate of 20 EUR, the cost would translate to 1.37 billion EUR. \\We recognize the invaluable role that this principle plays for the whole academia. Without changing this principle, we also recognize that refereeing is time-consuming and dedication to this service should be rewarded. We envision a list of incentives in exchange for well-done (i.e. without repeated delays, RR consistently different from CR, etc.) peer-reviewing. For instance, each refereeing job may grant the referee a number of points, which may be used for discounts on potential submission fees\footnote{\label{12}If they were needed to support the running costs of the infrastructure}, to grant a longer or shorter cool down time if desired, or to increase the visibility of the referee’s own works on the online portal, such as appearing in sections dedicated to higher ranking papers. 

\section{Final remarks}
The authors could still be charged a small sum to contribute to the maintenance of the servers and the portal. The charges will not be flat, but will consider the economic conditions of the country and institution of origin of the authors. For a list of countries and institutions, publication will be completely free or require only a symbolic contribution.\\
Lastly, we point out that many scientific journals have as prominent goal the dissemination of research to the lay public, and often offer a positive, high-profile contribution to the scientific issues of public interest, such as debating scientific regulation, funding, scientific policies, trends, accountability, etc. Journals will now be able to focus solely on this role, very much like the standard press has in the context of general information.

\begin{acknowledgments}
We wish to acknowledge the many colleagues, journal editors, and librarians we entertained endless discussions with over the past five years, while this project was slowly conceived, written, corrected, and put in a drawer until the next iteration. We mention in particular Johannes Fournier, Alessio Gagliardi, Marcus Huber, Johannes Kn\"{o}rzer, and Esther Tobschall.

\end{acknowledgments}

%\nocite{*}

\vfill\null

\end{document}